\documentclass[aps,preprint]{revtex4}%
\usepackage{amsfonts}
\usepackage{amsmath}
\usepackage{amssymb}
\usepackage{graphicx}%
\usepackage{hyperref}
\providecommand{\U}[1]{\protect\rule{.1in}{.1in}}
\begin{document}
 \pdfoutput=1
\preprint{HEP/123-qed}
\title[ ]{Einstein-Born-Infeld gravastar models, dark matter and accretion mechanisms}
\author{D.J. Cirilo-Lombardo}
\affiliation{CONICET- Universidad de Buenos Aires. Instituto de F\'\i sica del Plasma
(INFIP). Buenos Aires, Argentina}
\affiliation{Universidad de Buenos Aires. Facultad de Ciencias Exactas y Naturales.
Departamento de F\'\i sica. Buenos Aires, Argentina}
\affiliation{Bogoliubov Laboratory of Theoretical Physics, Joint Institute for Nuclear
Research, 141980 Dubna, Russian Federation}
\author{C.D. Vigh}
\affiliation{Instituto de Ciencias, Universidad Nacional de General Sarmiento, Los
Polvorines, Argentina}
\affiliation{Universidad de Buenos Aires. Facultad de Ciencias Exactas y Naturales.
Departamento de F\'\i sica. Buenos Aires, Argentina}
\affiliation{CONICET- Universidad de Buenos Aires. Instituto de F\'\i sica del Plasma
(INFIP). Buenos Aires, Argentina}
\pacs{PACS number}

\begin{abstract}
Gravastar models have recently been proposed as an alternative to black holes,
mainly to avoid the pro\-ble\-ma\-tic issues associated with event horizons
and singularities. In this work, a regular variety of gravastar models within
the context of Einstein-Born-Infeld (EBI) nonlinear electrodynamics are
builded. These models presented here are truly regular in the sense that both
the metric and its derivatives are continuous throughout space-time, contrary
to other cases in the literature where matching conditions are necessary in
the interior and exterior regions of the event horizon. We investigated the
accretion process for spherically symmetric space-time geometries generated
for a nonlinear electromagnetic field where the energy momentum tensor have
the same form that an anisotropic fluid that is the general EBI case. We
analyse this procedure using the most general static spherically symmetric
metric ansatz. In this theoretical context, we examined the accretion process
for specific sphe\-ri\-ca\-lly symmetric compact configuration obtaining the
accretion rates and the accretion velocities during the process and the flow
of the fluid around the black hole. In addition, we study the behaviour of the
rate of change of the mass for each chosen metric.

\end{abstract}
\volumeyear{year}
\volumenumber{number}
\issuenumber{number}
\eid{identifier}
\date[Date text]{date}
\received[Received text]{date}

\revised[Revised text]{date}

\accepted[Accepted text]{date}

\published[Published text]{date}

\startpage{101}
\endpage{102}
\maketitle
\tableofcontents

\section{Introduction}

The gravitational accretion of matter is ubiquitous in Astrophysics because it
is an efficient mechanism to convert gravitational energy into kinetic energy.
This kinetic energy could be converted into heat, radiation or power
relativistic jets \cite{gomez}. In essence, accretion is the process by which
a massive astrophysical object such as a black hole or a star can take
particles from a fluid from its vicinity which leads to increase in mass of
the accreting body \cite{BJ15}. Examples to the relevance of accretion
processes are involved with the existence of supermassive black holes at the
center of galaxies.

However, since the fact that black holes could be formed by gravitational
collapse of a massive star is not the only process. Another way could be the
merger of several small black holes in a specified domain where the conditions
are propicious but, its probability is too low \cite{BJ15}, \cite{AK02}.

Gravastar models have recently been proposed as an alternative to black holes,
mainly to avoid the pro\-ble\-ma\-tic issues associated with event horizons
and singularities \cite{MM04}. Gravastar: \textquotedblleft Gravitational
vacuum star\textquotedblright\ is a model in which a very small number of
serious challenges to our usual conception of a black hole \cite{VW04}. In the
original formulation by Mazur and Mottola, gravastars contain a central region
featuring a $p=-\rho$ false vacuum or "dark energy", a thin shell of $p=\rho$
perfect fluid, and a true vacuum $p=\rho=0$ exterior. The dark energy like
behavior of the inner region prevents collapse to a singularity and the presence
of the thin shell prevents the formation of an event horizon, avoiding the
infinite blue shift. The inner region has thermodynamically no entropy and may
be thought of as a gravitational Bose--Einstein condensate. Severe red-shifting
of photons as they climb out of the gravity well would make the fluid shell
also seem very cold, almost absolute zero. Several authors studied its
properties and stability conditions for the existence of such astrophysical
objects \cite{2006Deb}. As we know, it is generally well-accepted the notion of
a black hole in the general relativist community, but a considerable number of
theoretical particle and condensed matter physicists as esceptic with the
concept of event horizons and singularities of the spacetime. Also the existence
of misconceptions and the problems inherent to the Schwarzschild solution in
the literature put in more complicated form the black hole concept. The fact
that the evidence of the existence of black holes in the Universe is apparently
very convincing, several problems regarding the observational data is still
encountered. Consequently, this scepticism has inspired new ideas as models
replacing the interior Schwarzschild solution with compact structure in order
to deal with the problems of the singularity at the origin and the event
horizon. Then the interior structure of realistic black holes has not been
determined, and is still open to considerable debate. Some years ago, an
interesting alternative to black holes has been proposed: the \textquotedblleft
gravastar\textquotedblright\ picture developed by Mazur and Mottola \cite{MM04}.
In this model, the quantum vacuum undergoes a phase transition at or near the
location where the event horizon is formed. The Mazur-Mottola model is
constituted by an onion-like structure of a not so simple manner being the full
dynamic stability against spherically symmetric perturbations one the most
remarkable question. The other important question is the lack of the condition
of regularity in all solutions of the gravastar type inspired by the solution
of Mazur and Mottola. In \cite{MM04} was claimed that the solution is
thermodynamically stable, however due the structure of the model, other types of
analysis were di¢ cult to apply. The radial stability of a simplifed model with
three layers was performed in \cite{VW04} and it was found that only for a
number of configurations there was stability. The generalization was given in
\cite{car} for gravastar models with diferent exteriors. Other possibilities
that strongly motivate us to carry out the research developed in this paper were
the choice for the interior solution that have been considered in \cite{bilic}
replaced the standard de-Sitter interior by a Born-Infeld phantom, in
\cite{lobo06} the interior solution was changed by one that is governed by the
dark-energy equation of state (EOS), and in \cite{lobo07} an interior nonlinear
electrodynamic geometries were matched to the Schwarzschild exterior. The other
important question is the lack of the condition of regularity in all solutions
of the gravastar type inspired by the solution of Mazur and Mottola. Other very
interesting proposals as gravastar models beside the \cite{MM04} and the
3-layer model (infinitesimally thin shell of matter) of \cite{VW04} are the
fluid gravastar model (no shells) \cite{catto}, electrically charged gravastars
\cite{dubra}, magnetized gravastars \cite{turi} and rotating gravastars
\cite{nami}. All these proposals are conditioned to meet the corresponding
standard characteristics of gravastar with respect to black holes, such as low
entropy (which would correspond to a Bose condensate as the final state of a
star).

Several authors investigated more recently how gravastar theory is consistent
with real observational data. Evaluating stability properties \cite{ceci} and
looking for that if is possible to distinguish observational signatures like
energy flux, temperature distribution and equilibrium radiation spectrum
between gravastars and black holes \cite{harko}. More specifically, comparing
the energy flux emerging from the surface of the thin accretion disk around
black holes and gravastars of similar masses, it was found that its maximal
value is systematically lower for gravastars, independently of the values of
the spin parameter or the quadrupole momentum. These effects are confirmed from
the analysis of the disk temperatures and disk spectra. In addition to this, it
is also shown that the conversion efficiency of the accreting mass into
radiation is always smaller than the conversion efficiency for black holes,
i.e., gravastars provide a less efficient mechanism for converting mass to
radiation than black holes. However, the discussion is still open \cite{natr}
being precisely the issue of efficiency one of our motivations, due to the
nature of the model that we will present in this paper with respect to the
traditional concept of gravastar.

In this work, an alternative regular variety of gravastar models within the
context of EBI nonlinear electrodynamics are builded. We investigated the
accretion process for spherically symmetric space-time geometries generated
for a nonlinear electromagnetic field where the energy momentum tensor has the
same form that an anisotropic fluid that is the case of the EBI model. We have
analyzed this procedure using the most general nonlinear black hole metric
solution in the EBI model. We study the gravitational vacuum star (gravastar)
configuration as proposed by other authors in a model where the regular interior
(similar to the "compact" de Sitter spacetime segment given by other authors) is
continuously extended to the exterior spacetime that approaching
Schwarzchild-Reissner Nordstrom at asymptotic distances. Consequently the
multilayered structure in previous references is replaced by a continuous
stress-energy tensor as in the case of reference \cite{VW04} (given in other
theoretical context) were the proposed metrics were constructed by hand instead
from first principles as here.

In particular(to analize the accretion process) we are going to follow a
parallel way that it was employed by Bahamonde et al (2015) \cite{BJ15}, using
the metric studied in Cirilo-Lombardo (2005) \cite{DCL} and a new one
presented here of the Yukawa (e.g. exponential, fifth force) type. Combining
these results we are going to study and analyze accretion rate, accretion
velocity and the regularity of the proposed metric both: without matter and
with matter (e.g.: exotic and normal one). In the next two sections, before to
enter in the accretion problem, we review our previous results concerning
EBI\ regular solutions to be this work self-contained without enter in full
details (see \cite{DCL}\cite{di6})

\section{Born Infeld theory}

In 1934 M. Born and L. Infeld \cite{bi} introduced the most relevant version
of the non-linear electrodynamics with, among others, these main properties:

\textit{\ i) Geometrically} the Born-Infeld (BI) Lagrangian density is one of
the most simplest non-polinomial Lagrangian densities that is invariant under
the general coordinate transformations.

ii) The BI electrodynamics is the only \textit{causal} spin-1 theory
\cite{des}\cite{tt}aside the Maxwell theory. The vacuum is characterized with
$F_{\mu\nu}=0$ and the \textit{energy density} is definite semi-positive.

iii) The BI theory \textit{conserves helicity }\cite{ish} and solves the
problem of the \textit{self-energy} of the charged particles \cite{bi}%
\cite{ti}.

The Lagrangian density describing Born-Infeld theory (in arbitrary spacetime
dimensions) is
\begin{equation}
\mathcal{L}_{BI}=\sqrt{-g}L_{BI}=\frac{b^{2}}{4\pi}\left\{  \sqrt{-g}%
-\sqrt{\left\vert \det(g_{\mu\nu}+b^{-1}F_{\mu\nu})\right\vert }\right\}
\label{la}%
\end{equation}
where $b$ is a fundamental parameter of the theory with field dimensions. In
open superstring theory \cite{ss}, for example, loop calculations lead to
this Lagrangian with $b^{-1}=2\pi\alpha^{\prime}$ ($\alpha^{\prime}\equiv
\ $inverse of the string tension)\ . In four spacetime dimensions the
determinant in (\ref{la}) may be expanded out to give
\begin{equation}
L_{BI}=\frac{b^{2}}{4\pi}\left\{  1-\sqrt{1+\frac{1}{2}b^{-2}F_{\mu\nu}%
F^{\mu\nu}-\frac{1}{16}b^{-4}\left(  F_{\mu\nu}\widetilde{F}^{\mu\nu}\right)
^{2}}\right\}
\end{equation}
(were $F^{\mu\nu}$ is the electromagnetic field and its dual defined as
$\widetilde{F}^{\mu\nu}=\frac{1}{2}\varepsilon^{\mu\nu\gamma\delta}%
F_{\gamma\delta})$ which coincides with the usual Maxwell Lagrangian in the
weak field limit.

Recently, interest has been rising in this non-linear electromagnetic theory
since it has turned out to play an important role in the development of the
string theory, as was very well described in the pioneering work of
\ Barbashov and Chernikov \cite{bc1}, \cite{bc2}. The non-linear
electrodynamics of the Born-Infeld Lagrangian, shown in \cite{az}, describes
the low energy process on D-branes which are non-perturbative solitonic objects
that arises for the natural D-dimensional extension of the string theory. The
structure of the string theory was improved significantly with the introduction
of the D-branes, because many physically realistic models can be constructed.
For example, the well known \textquotedblright brane-world\textquotedblright%
\ scenario that naturally introduces the BI electrodynamics into the gauge
theories.

Another interesting feature is with the recent advent of the physics of
D-branes, the solitons in the non-perturbative spectrum of string theory, it
has been realized that their low energy-dynamics can be properly described by
the so called Dirac-Born-Infeld (DBI) action \cite{le}\cite{az}. Since single
branes are known to be described by the abelian DBI action, one migth expect
naturally that multiple brane configurations would be a non abelian
generalization of the Born-Infeld action. Specifically in the case of
superstring theory one has to deal with a supersymmetric extension of DBI
actions and when the number of D-branes coincides there is a symmetry
enhancement \cite{wi} and the abelian DBI\ action should be generalized to its
non abelian counterpart being the more consistent development proposed for
non-abelian supersymmetric extension was given in ref.\cite{di6}

\section{Class of EBI\ regular metrics}

In reference\cite{DCL}, the completely regular solution of the Born-Infeld
model coupled to gravity was found. The starting point was the general line
element given by%
\begin{equation}
ds^{2}=-e^{2\Lambda}dt^{2}+e^{2\Phi}dr^{2}+e^{2F\left(  r\right)  }d\theta
^{2}+e^{2G\left(  r\right)  }\sin^{2}\theta\,d\varphi^{2} \label{e}%
\end{equation}
Notice that above form of the metric ansatz with corresponds to an anisotropic
fluid because the EBI\ energy-momentum tensor have such symmetry. The general
solution of the system of EBI\ equations take of the following form%
\begin{equation}
ds^{2}=-e^{2\Lambda}dt^{2}+e^{2\mathcal{F}\left(  r\right)  }\left[
e^{-2\Lambda}\left(  1+r\,\ \partial_{r}\mathcal{F}\left(  r\right)
\,\right)  ^{2}dr^{2}+r^{2}\left(  d\theta^{2}+\sin^{2}\theta\,d\varphi
^{2}\right)  \right]  \label{int}%
\end{equation}
The function $\mathcal{F}\left(  r\right)  $ is determined in such a way that
the electric field of the configuration obey the following requirements to
give a regular solution in the sense that was given by B. Hoffmann and L.
Infeld(see with detail in\cite{DCL})
\begin{equation}
\left.  F_{01}\right\vert _{r=r_{o}}<b
\end{equation}%
\begin{equation}
\left.  F_{01}\right\vert _{r=0}=0
\end{equation}%
\begin{equation}
\,\ \ \ \ \ \ \ \ \ \ \ \ \ \ \ \ \ \ \ \ \ \ \ \ \ \ \left.  F_{01}%
\right\vert _{r\rightarrow\infty}%
=0\,\ \ \ \ \ \ \ \ \ \ \ \ \text{assymptotically \ Coulomb}%
\end{equation}
\ and is directly associated to the Kernel function $K_{i}\left(  r\right)  $
that will be defined in Section IV. The metric solution is based on the
observation that the energy momentum tensor of EBI can be written as a
relativistic fluid with $\left\vert \rho\right\vert =\left\vert p\right\vert $
(where in this case density and pressure are functions of the electromagnetic
fields), and consequently, the general ansatz must be of the form
(\ref{action}).The substitution
\begin{equation}
Y\equiv r\,e^{\mathcal{F}\left(  r\right)  }%
\end{equation}
and differentiating it
\begin{equation}
dY\equiv\,e^{\mathcal{F}\left(  r\right)  }\left(  1+r\,\ \partial
_{r}\mathcal{F}\left(  r\right)  \,\right)  dr
\end{equation}
the interval (\ref{int}) takes the form
\begin{equation}
ds^{2}=-e^{2\Lambda}dt^{2}+e^{-2\Lambda}dY^{2}+Y^{2}\left(  d\theta^{2}%
+\sin^{2}\theta\,d\varphi^{2}\right)
\end{equation}
we can see that the metric (in particular the $g_{tt}$ coefficient) takes the
similar form like a Demianski solution for the Born-Infeld monopole spacetime
:
\begin{equation}
e^{2\Lambda}=1-\frac{2M}{Y}-\frac{2b^{2}r_{o}^{4}}{3\left(  \sqrt{Y^{4}%
+r_{o}^{4}}+Y^{2}\right)  }-\frac{4}{3}b^{2}r_{o}^{2}\,_{2}F_{1}\left[
1/4,1/2,5/4;-\left(  \frac{Y}{r_{0}}\right)  ^{4}\right]
\end{equation}
here $M$ is an integration constant, which can be interpreted as an intrinsic
mass, and $_{2}F_{1}$ is the Gauss hypergeometric function. We have pass
\begin{equation}
g_{rr}\rightarrow g_{YY},\,\ \ \ \ \ \ \ \ \ \ \ \ \ \ g_{tt}\left(  r\right)
\rightarrow g_{tt}\left(  Y\right)
\end{equation}
Specifically, for the form of the $\mathcal{F}\left(  r\right)  $
from\cite{DCL}, $Y$ is
\begin{equation}
Y^{2}\equiv\left[  1-\left(  \frac{r_{o}}{a\left\vert r\right\vert }\right)
^{n}\right]  ^{2m}r^{2}%
\end{equation}
Now, with the metric coefficients fixed to a asymptotically Minkowskian form,
one can study the asymptotic behaviour of our solution. A regular,
asymptotically flat solution with the electric field and energy-momentum
tensor both regular, in the sense of B. Hoffmann and L. Infeld is when the
exponent numbers of $Y(r)$ take the following particular values:
\begin{equation}
n=3\ \ \ and\ \ \ \ \ m=1
\end{equation}
In this case, and for $r>>\frac{r_{0}}{a}$ $,$ we have the following
asymptotic behaviour for $Y\left(  r\right)  $ and $-g_{tt}\,$, that does not
depend on the $a$ parameter%
\begin{equation}
Y\left(  r\right)  \rightarrow r\ \text{ \ \ \ \ \ }\left(  r>>\frac{r_{0}}%
{a}\right)
\end{equation}%
\begin{equation}
e^{2\Lambda}\simeq1-\frac{2M}{r}-\frac{8b^{2}r_{o}^{4}K\left(  1/2\right)
}{3r_{o}r}+2\frac{b^{2}r_{o}^{4}}{r^{2}}+...
\end{equation}
A distant observer will associate with this solution a total mass
\begin{equation}
M_{eff}=M+\frac{4b^{2}r_{o}^{4}K\left(  1/2\right)  }{3r_{o}}%
\end{equation}
here $K\left(  m\right)  \,$\ complete elliptic function of first kind, and
total charge
\begin{equation}
Q^{2}=2b^{2}r_{o}^{2}%
\end{equation}
Notice that when the intrinsic (Schwarzchild) mass $M$ is zero the line
element is regular everywhere, the Riemann tensor is also regular everywhere
and hence the space-time is singularity free. The electromagnetic mass
\begin{equation}
M_{el}=\frac{4b^{2}r_{o}^{4}K\left(  1/2\right)  }{3r_{o}} \label{mel}%
\end{equation}
and the charge $Q$ are the \textit{twice} that the electromagnetic charge and
mass of the Demianski solution \cite{dem} for the static electromagnetic geon.
Notice that the $M_{el}$ is necessarily positive, which was not the case in
the Schwarzschild line element. The other important reason for to take the
constant $M=0$ is that we must regard the quantity (let us to restore by one
moment the gravitational constant $G$)
\begin{equation}
4\pi G\int_{Y(r=0)}^{Y(r)}T_{0}^{0}\left(  Y\right)  Y^{2}dY
\end{equation}
as the \textit{gravitational mass} causing the field at coordinate distance r
from the pole. In our case $T_{0}^{0}$ is precisely (in gravitational units)
$M_{el}$ given by (\ref{mel}), the \textit{total electromagnetic mass} within
the sphere having its center at $r=0$ and coordinate $r.$ We will take $M=0$
in the rest of the analysis of this paragraph only.

On the other hand, the function $Y\left(  r\right)  $ for the values of the
$m$ and $n$ parameters given above has the following behaviour near of the
origin
\[
\text{for\ }a<0\text{ \ \ \ \ \ \ \ \ when\ }r\rightarrow0,\ Y\left(
r\right)  \rightarrow\infty
\]%
\[
\text{for\ }a>0\text{ \ \ \ \ \ \ \ \ \ when \ }r\rightarrow0,\ Y\left(
r\right)  \rightarrow-\infty
\]
Notice that the case \ $a>0$ will be excluded because in any value
$r_{0}\rightarrow$ $Y\left(  r_{0}\right)  =0$ , the electric field takes the
limit value $b$ and the regularity condition is violated. For $M=0$ and
$a<0,$expanding the hypergeometric function, we can see that the $-g_{tt}$
coefficient has the following behaviour near the origin
\begin{equation}
e^{2\Lambda}\simeq1-\frac{8b^{2}r_{o}^{4}K\left(  1/2\right)  }{3r_{o}}%
r^{2}\left(  \frac{\left\vert a\right\vert }{r_{0}}\right)  ^{3}+2b^{2}%
r_{0}^{4}\ r^{4}\left(  \frac{\left\vert a\right\vert }{r_{0}}\right)
^{6}+...
\end{equation}
The metric (see figures below) and the energy-momentum tensor remains
\textit{both }regulars at the origin (it is: $g_{tt}\rightarrow-1,T_{\mu\nu
}\rightarrow0$ $\ $\ for $r\rightarrow0$). It is not very difficult to check
that (for $m=1$ and $n=3$) the maximum of the electric field (see figures
below) is not in $r=0$ , but in the \textit{physical} \textit{border} of the
spherical configuration source of the electromagnetic fields (this point is
located around $r_{B}=2^{1/3}\frac{r_{0}}{\left\vert a\right\vert }$). It
means that $Y\left(  r\right)  $ maps correctly the internal structure of the
source in the same form that the quasi-global coordinate of the
reference \cite{br} for the global monopole in general relativity. The lack of
the conical singularities at the origin is because the very well description
of the manifold in the neighborhood of $r=0$ given by $Y\left(  r\right)  $

Because the metric is regular ($g_{tt}=-1,$ at $r=0$ and at $r=\infty$), its
derivative must change sign. In the usual gravitational theory of general
relativity the derivative of $g_{tt}$ is proportional to the gravitational
force which would act on a test particle in the Newtonian approximation. In
Einstein-Born-Infeld theory with this new static solution, it is interesting
to note that although this force is attractive for distances of the order
$r_{0}<<r$ , it is actually a repulsion for very small $r$. For $r$ greater
than $r_{0},$ the line element closely approximates to the Schwarzschild form:
consequently can simplify the association with a continuos \textit{gravastar
model without matching conditions}. Thus the regularity condition shows that
the electromagnetic and gravitational mass are the same and, as in the
Newtonian theory, we now have the result that the attraction is zero in the
center of the spherical configuration source of the electromagnetic field.

The next figures\ref{1} and \ref{2} show the electric field $F_{10}$ (tetrad
coordinates from \cite{DCL})of the EBI - monopole in function of r,
for$M=0,r_{0}=1,m=1,n=3$ and a=-0.9 and the coefficient $-g_{tt}$of the EBI -
monopole as a function of r, for$M=0,r_{0}=1,m=1,$ n=3 and a=-0.9. Notice that
at once the singularity of the electric field is solved, the resulting metric
is automatically without singularities as is clear in the context of the
general relativity due the system equation of EBI. \
%TCIMACRO{\FRAME{ftbpFU}{5.8902in}{3.6296in}{0pt}{\Qcb{{}Electric field
%$F_{10}$ of the EBI- monopole as a function of r, for $M=0$, $r_{0}=1$, $m=1$,
%$n=3$ and $a=-0.9$ (see deatails in \cite{DCL})}}{\Qlb{1}}{Figure}%
%maintain-aspect-ratio TRUE;  display "USEDEF";  valid_file "T";
%width 5.8902in;  height 3.6296in;  depth 0pt;  original-width 7.1684in;
%original-height 4.4053in;  cropleft "0";  croptop "1";  cropright "1";
%cropbottom "0";  tempfilename 'PHGU8F00.wmf';tempfile-properties "XPR";}} }%
%BeginExpansion
\begin{figure}
[ptb]
\begin{center}
\includegraphics[
%natheight=4.405300in,
%natwidth=7.168400in,
%height=3.6296in,
%width=5.8902in
scale=.5]{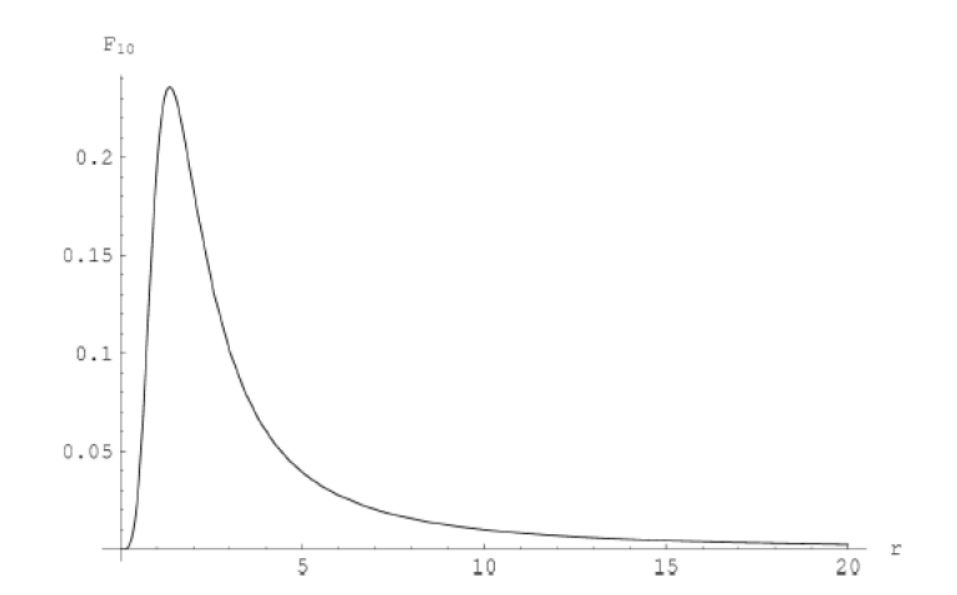}%
%{PHGU8F00.wmf}%
\caption{{}Electric field $F_{10}$ of the EBI- monopole as a function of r,
for $M=0$, $r_{0}=1$, $m=1$, $n=3$ and $a=-0.9$ (see deatails in \cite{DCL})}%
\label{1}%
\end{center}
\end{figure}
%EndExpansion
%TCIMACRO{\FRAME{ftbpFU}{6.1704in}{3.6115in}{0pt}{\Qcb{{}Coefficient $-gtt$ of
%the EBI - monopole in function of r, for$M=0$, $r_{0}=1$, $m=1$, $n=3$ and
%$a=-0.9$ (see details in \cite{DCL})}}{\Qlb{2}}{Figure}%
%maintain-aspect-ratio TRUE;  display "USEDEF";  valid_file "T";
%width 6.1704in;  height 3.6115in;  depth 0pt;  original-width 8.6749in;
%original-height 5.0626in;  cropleft "0";  croptop "1";  cropright "1.0002";
%cropbottom "0";  tempfilename 'PHGWFM04.wmf';tempfile-properties "XPR";}} }%
%BeginExpansion
\begin{figure}
[ptb]
\begin{center}
\includegraphics[
%trim=0.000000in 0.000000in -0.001735in 0.000000in,
%natheight=5.062600in,
%natwidth=8.674900in,
%height=3.6115in,
%width=6.1704in
scale=.5]{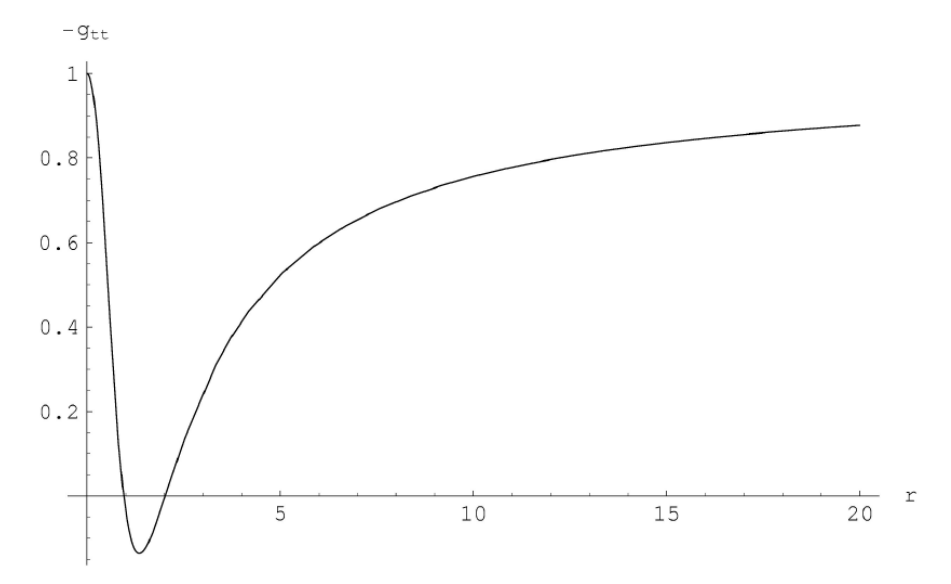}%
%{PHGWFM04.wmf}%
\caption{{}Coefficient $-gtt$ of the EBI - monopole in function of r,
for$M=0$, $r_{0}=1$, $m=1$, $n=3$ and $a=-0.9$ (see details in \cite{DCL})}%
\label{2}%
\end{center}
\end{figure}
%EndExpansion

\subsection{Equations for the electromagnetic fields}

The equations that describe the dynamic of the electromagnetic fields of
Born-Infeld in a curved spacetime are
\begin{equation}
\nabla_{a}\mathbb{F}^{ab}=\nabla_{a}\left[  \frac{F^{ab}}{\mathbb{R}}+\frac
{P}{b^{2}\mathbb{R}}\widetilde{F}^{ab}\right]
=0\,\ \ \ \ \ \ \ \ \ \ \ \ \ \ \ \ \ \ \ \ \ \ \left(
field\,equations\,\right)
\end{equation}%
\begin{equation}
\nabla_{a}\,\,\widetilde{F}^{ab}%
=0\,\ \ \ \ \ \ \ \ \ \ \ \ \ \ \ \ \ \ \ \ \ \ \ \ \ \ \ \ \ (\ Bianchi^{\prime
}%
s\,\ identity)\ \ \ \ \ \ \ \ \ \ \ \ \ \ \ \ \ \ \ \ \ \ \ \ \ \ \ \ \ \ \ \ \ \
\end{equation}
where the scalar and pseudoscalar invariants are
\begin{equation}
P\equiv-\frac{1}{4}F_{\alpha\beta}\widetilde{F}^{\alpha\beta}%
\end{equation}%
\begin{equation}
S\equiv-\frac{1}{4}F_{\alpha\beta}F^{\alpha\beta}%
\end{equation}%
\begin{equation}
\mathbb{R}\equiv\sqrt{1-\frac{2S}{b^{2}}-\left(  \frac{P}{b^{2}}\right)  ^{2}}%
\end{equation}
The above equations can be solved explicitly in the tetrad from the line
element(\ref{e})giving the follow result
\begin{equation}
F_{01}=A\left(  r\right)
\end{equation}

\begin{equation}
\mathbb{F}_{01}=f\ e^{-2G}%
\end{equation}
where $f$ is a constant. We can see from the above eqs. that
\[
\mathbb{F}_{01}=\frac{F_{01}}{\sqrt{1-\left(  \overline{F}_{01}\right)  ^{2}}}%
\]
where we obtain the following form for the electric field of the
self-gravitating BI monopole
\begin{equation}
F_{01}=\frac{b}{\sqrt{\left(  \frac{b}{f}e^{2G}\right)  ^{2}+1}}%
\end{equation}
Consequently, having into account the explicit expression for $f$%
\begin{equation}
f=br_{0}^{2}\equiv Q\,\ \ \ \ \Rightarrow\,\ \ \ F_{01}=\frac{b}{\sqrt{\left(
\frac{e^{G}}{r_{0}}\right)  ^{4}+1}}%
\end{equation}
The magnetic field is solved in an analog manner considering $F_{23}$ instead
$F_{10}$ \cite{DCL}. Where $r_{0}$ is a constant with units of longitude that in
reference \cite{bi} was associated to the radius of the electron. Finally the
components of the energy-momentum tensor of BI takes its explicit form using
the $F_{01}$ that we was found, namely
\begin{equation}
-T_{00}=T_{11}=\frac{b^{2}}{4\pi}\left(  1-\sqrt{\left(  \frac{r_{0}}{e^{G}%
}\right)  ^{4}+1}\right)
\end{equation}%
\begin{equation}
T_{22}=T_{33}=\frac{b^{2}}{4\pi}\left(  1-\frac{1}{\sqrt{\left(  \frac{r_{0}%
}{e^{G}}\right)  ^{4}+1}}\right)
\end{equation}

\section{General formalism}

The starting point is the solution reviewed in the previous sections, that we
rewrite as follows
\begin{equation}
ds^{2}=-A(r)dt^{2}+\frac{1}{B(r)}dr^{2}+C(r)(d\theta^{2}+\sin^{2}\theta
d\phi^{2}) \label{action}%
\end{equation}
(e.g.: general static spherically symmetric space-time supposed with
stationary regime) where the functions $A(r),B(r)$ and $C(r)$ have been
determined according the general metric solution in Cirilo-Lombardo (2005)
\cite{DCL}:
\begin{equation}
A(r)=e^{2\Lambda}%
\end{equation}%
\begin{equation}
B(r)=A(r)\left[  \frac{dY_{i}(r)}{dr}\right]  ^{-2}%
\end{equation}
%B(r)=e^{2\Lambda}\left[  1-\left(  \frac{r_{0}}{a|r|}\right)  ^{n}\right]
%^{2(1-m)}\left[  1-\left(  \frac{r_{0}}{a|r|}\right)  ^{n}(m~n-1)\right]
%^{-2}%
\begin{equation}
C(r)=r^{2}\left[  K_{i}(r)\right]  ^{2m}%
\end{equation}%
\begin{equation}
e^{2\Lambda}=1-\frac{2M}{Y_{i}}-\frac{2b^{2}r_{0}^{4}}{3(Y_{i}^{2}+\sqrt
{Y_{i}^{4}+r_{0}^{4}})}-\frac{4}{3}b^{2}r_{0}^{2}~~_{2}F_{1}%
[1/4,1/2,5/4,-(Y/r_{0})^{4}]
\end{equation}
where
\begin{equation}
r_{0}=\sqrt{\frac{e}{b}}%
\end{equation}
$e$ is the elementary charge and $b$ the fundamental field strength.
$_{2}F_{1}[1/4,1/2,5/4,-(Y(r)/r_{0})^{4}]$ is the standard Hypergeometric
Gauss Function and
\begin{equation}
Y_{i}(r)\equiv r\left[  K_{i}(r)\right]  ^{m}\qquad i=1,2
\end{equation}
Where we defined
\begin{equation}
K_{1}(r)\equiv1-\left(  \frac{r_{0}}{a|r|}\right)  ^{n}%
\end{equation}%
\begin{equation}
K_{2}(r)\equiv\exp\left\{  -\frac{r_{0}}{a|r|}n\right\}
\end{equation}
being $K_{1}$ the "kernel" of $Y_{1}$ proposed in Cirilo-Lombardo (2005)
\cite{DCL}and $K_{2}$ the new one proposed in this work for $Y_{2}$. Note that
clearly $K_{1}$ and $K_{2}$ are subject to the properties of the
electromagnetic field (due the boundary conditions, see Section II) by means
of the BI energy-moment tensor, consequently modifying the metric
(specifically coefficient A, B and C see\ref{action}) through the
gravitational EBI\ equations (they do not constitute a gauge due the boundary
conditions as described in the Introduction). The advantage of $K_{2}(r)$over
$K_{1}(r)$ is that the cusps of conical character in all the curves for n=1
near the origin are eliminated due the exponential (Yukawa) behaviour. Also
the asymptotic behaviour is improved and controlled for all parameters of the
model. Note that, as we saw earlier in $\mathcal{F}$, $K_{i}(r)$ is arbitrary
but subject to the boundary conditions fixed by the physical scenario and the
regularity conditions in the sense of \cite{DCL}

\section{Dynamic equations of motion}

In this section, we consider the simplest case of spherically symmetric
stationary accretion. The stationarity assumes that the BH mass increases
slowly, such that the distribution of the fluid on the relevant space-time
scales has time to adjust itself to the changing BH metric. Following the
procedure in Bahamonde and Jamil \cite{BJ15}we can write the 4-velocity under
the restriction of spherical symmetry and stationary regime:
\begin{equation}
u^{\mu}=\frac{dx^{\mu}}{d\tau}=(u^{t},u^{r},0,0)
\end{equation}
where $\tau$ is the proper time. This is a necessary step to compute the
accretion rate. In this case all the variables are functions of $r$. Imposing
the normalization condition, $u_{\mu}u^{\mu}=-1$ we have:\newline%
\begin{equation}
u^{t}\equiv\frac{dt}{d\tau}=\sqrt{\frac{u^{2}+B}{AB}}%
\end{equation}%
\begin{equation}
u^{r}=\frac{dr}{dt}%
\end{equation}
From the energy-momentum conservation law:
\begin{equation}
0=T_{;\mu}^{\mu\nu}=\frac{1}{\sqrt{-g}}(\sqrt{-g})T_{,\mu}^{\mu\nu}%
+\Gamma_{\alpha\mu}^{\nu}T^{\alpha\mu}%
\end{equation}
we find:
\begin{equation}
(\rho+p)uC(r)\frac{A(r)}{B(r)}\sqrt{u^{2}+B(r)}=A_{1} \label{a1}%
\end{equation}
\newline Projecting the 4-velocity onto the energy momentum conservation law
we can obtain the energy flux continuity equation:
\begin{equation}
u^{\mu}\rho_{,\mu}+(\rho+p)u_{;\mu}^{\mu}=0 \label{u}%
\end{equation}
Pressure $p$ and density $\rho$ are related by an equation of state as:
\begin{equation}
p=p(\rho)
\end{equation}
It is possible to obtain from (\ref{u}):
\begin{equation}
\frac{\rho^{\prime}}{\rho+p}+\frac{u^{\prime}}{u}+\frac{A^{\prime}}{2A}%
+\frac{B^{\prime}}{2B}+\frac{C^{\prime}}{C}=0
\end{equation}
where primes denotes derivatives respect to $r$. If we integrate this:
\begin{equation}
uC(r)\sqrt{\frac{A(r)}{B(r)}}e^{\int\frac{d\rho}{\rho+p(\rho)}}=-A_{0}%
\end{equation}
If we study accretion we need $u$; due to $A_{0}$ is an integration constant
and, combining the fact that $T_{;\mu}^{\mu\nu}=0$
\begin{equation}
(\rho+p)\sqrt{u^{2}+B(r)}\sqrt{\frac{A(r)}{B(r)}}e^{-\int\frac{d\rho}%
{\rho+p(\rho)}}=A_{3} \label{a3}%
\end{equation}
where we defined $A_{3}=-A_{1}/A_{0}$. For spherical symmetry we take
$\theta=\pi/2$ and we obtain from the metric determinant $\sqrt{-g}%
=C\sqrt{A/B}$. Finally, considering the equation of mass flux, $0=J_{;\mu
}^{\mu}=\frac{1}{\sqrt{-g}}\frac{d}{dr}(J^{r}\sqrt{-g})$, we have:
\begin{equation}
\rho uC(r)\sqrt{\frac{A(r)}{B(r)}}=A_{2} \label{a2}%
\end{equation}
If we divide equations (\ref{a1}) and (\ref{a2}) we have:
\begin{equation}
\frac{(\rho+p)}{\rho}\sqrt{\frac{A(r)}{B(r)}}\sqrt{u^{2}+B(r)}=\frac{A_{1}%
}{A_{2}}\equiv A_{4} \label{a4}%
\end{equation}
$A_{1}$, $A_{2}$, $A_{3}$ and $A_{4}$ are integration constants. From the
previous expression the accretion velocity is straighforward deduced. If we
differentiate (\ref{a2}) and (\ref{a4}) following the procedure of
\cite{BJ15}, we can obtain, the accretion rate:
\begin{equation}
\dot{M}_{acc}=4\pi A_{3}M_{eff}^{2}[p(r)+\rho(r)]
\end{equation}
where the effective mass is:
\begin{equation}
M_{eff}\equiv2M+Y_{i}(r)\left(  \frac{2br_{0}^{4}}{3Y_{i}^{2}(r)+3\sqrt
{Y_{i}^{4}(r)+r_{0}^{4}}}-\frac{4}{3}b^{2}r_{0}^{2}H(r)\right)
\end{equation}
Due the absolute regular solutions (in the sense of \citep{DCL}), the density
electromagnetically induced appearing in is:
\begin{equation}
\rho(r)=\frac{b^{4}r_{0}^{6}}{4\pi Y_{i}^{2}(r)}(\sqrt{1+Y_{i}^{4}(r)}%
-Y_{i}^{2}(r))
\end{equation}
and the simplest choice for state equation for pressure, in order to compare
with Bahamonde and Jamil's paper \cite{BJ15}, is:
\begin{equation}
p(r)=(1+\omega)\rho(r)
\end{equation}
Where $\omega$ is a correction(constant shift) to include different kinds of
matter. Note that it is usually assumed in astrophysical scenarios that the
matter flow accompanies the electromagnetic field lines. Finally, the
expressions for electric $(F_{01})$and magnetic fields $(F_{23})$ are
respectively:
\begin{equation}
E(r)=\frac{b}{\sqrt{1+Y_{i}^{4}(r)}}%
\end{equation}%
\begin{equation}
B(r)=\frac{b}{Y_{i}^{2}(r)}%
\end{equation}

\subsection{Particular cases for $m=1$, $n=\{1,3,5,7\}$ and $\omega=0$}

\begin{figure}
[ptb]
\begin{center}
\includegraphics[
%natheight=3.133200in,
%natwidth=9.032100in,
%height=1.8758in,
%width=5.38in
scale=.5]{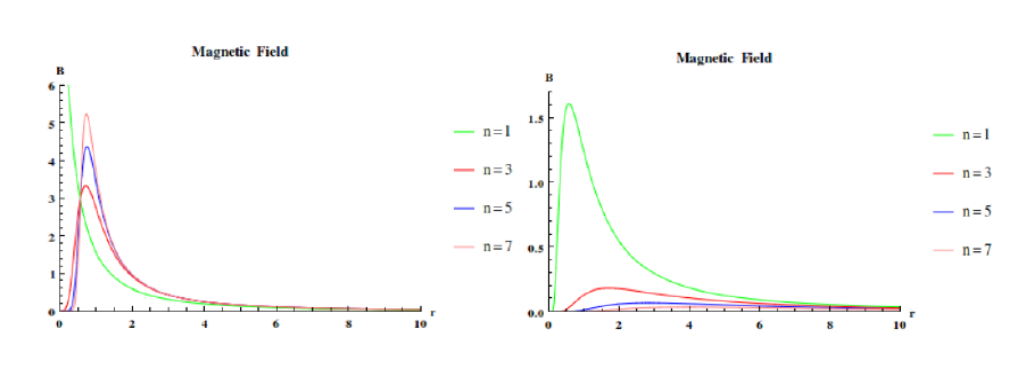}%
%{POIP5O04.wmf}%
\caption{Magnetic field for $n=1,3,5,7$. (a) \textbf{Left}: Using $K_{1}$. (b)
\textbf{Right}: Using $K_{2}$.}%
\label{103}%
\end{center}
\end{figure}

\begin{figure}
[ptb]
\begin{center}
\includegraphics[
%natheight=2.843500in,
%natwidth=8.794300in,
%height=1.9597in,
%width=6.026in
scale=.4]{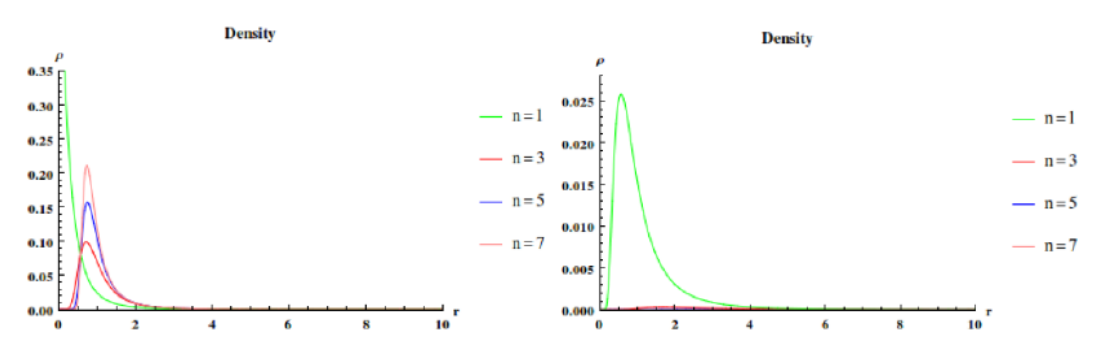}%
%{POIP8H05.wmf}%
\caption{Density for $n=1,3,5,7$. (a) \textbf{Left}: Using $K_{1}$. (b)
\textbf{Right}: Using $K_{2}$.}%
\label{104}%
\end{center}
\end{figure}

\begin{figure}
[ptb]
\begin{center}
\includegraphics[
%natheight=2.804600in,
%natwidth=8.873000in,
%height=1.7772in,
%width=5.585in
scale=.5]{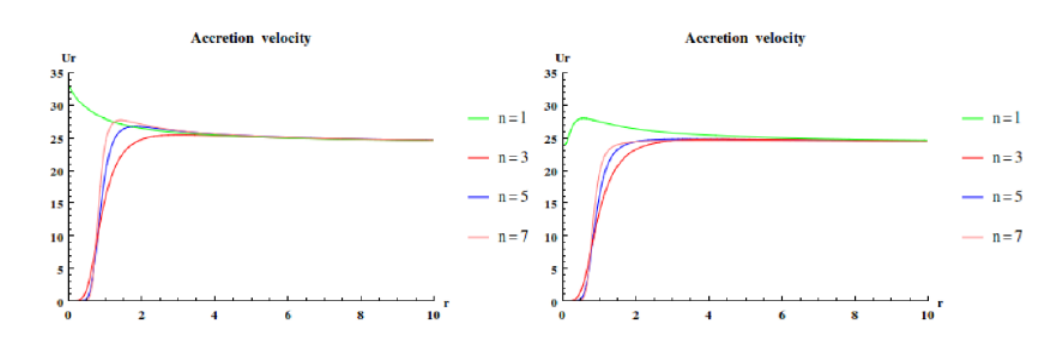}%
%{POIPA706.wmf}%
\caption{Accretion velocity for $n=1,3,5,7$. (a) \textbf{Left}: Using $K_{1}$.
(b) \textbf{Right}: Using $K_{2}$.}%
\label{105}%
\end{center}
\end{figure}

In this section we had set for some selected values of $m$ and $n$ and
computed the profiles of the metric coefficient, electric and magnetic fields,
accretion velocity, accretion mass rate and density. In order to fix the main
ideas, we had chosen the following values: $A_{4}=10$, $A_{2}=1$, $m=1$,
$a=-0.9$ $\ $and normalized ones $M=1$, $e=1$, $b=3.8296591$ for all profiles.%
%TCIMACRO{\FRAME{ftbpFU}{4.9199in}{4.267in}{0pt}{\Qcb{Metric coefficient $A(r)$
%for $n=1,3,5,7$. (a) \QTR{bf}{Top}: Using $K_{1}$. (b) \QTR{bf}{Bottom}: Using
%$K_{2\text{.}}$}}{\Qlb{101}}{Figure}{\special{ language "Scientific Word";
%type "GRAPHIC";  maintain-aspect-ratio TRUE;  display "USEDEF";
%valid_file "T";  width 4.9199in;  height 4.267in;  depth 0pt;
%original-width 2.6074in;  original-height 2.258in;  cropleft "0";
%tempfilename 'POIOX701.wmf';tempfile-properties "XPR";}} }%
%BeginExpansion
\begin{figure}
[ptb]
\begin{center}
\includegraphics[
%natheight=2.258000in,
%natwidth=2.607400in,
%height=4.267in,
%width=4.9199in
scale=.5]{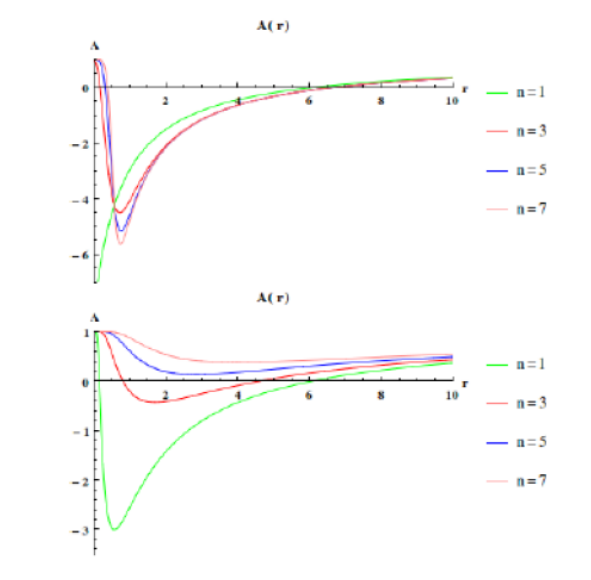}%
%{POIOX701.wmf}%
\caption{Metric coefficient $A(r)$ for $n=1,3,5,7$. (a) \textbf{Top}: Using
$K_{1}$. (b) \textbf{Bottom}: Using $K_{2\text{.}}$}%
\label{101}%
\end{center}
\end{figure}
%EndExpansion
In Fig.\ref{101} we showed the profile of the metric coefficient for both
kernels, $K_{1}$ at the top and $K_{2}$ at the bottom. The most clear effect
imposed by $K_{2}$ is smoothing for $n={1,3,5,7}$ and regularized the profile
of $A(r)$ for $n=1$.%
%TCIMACRO{\FRAME{ftbpFU}{7.1018in}{2.2208in}{0pt}{\Qcb{Electric field for
%$n=1,3,5,7$. (a) \QTR{bf}{Left}: Using $K_{1}$. (b) \QTR{bf}{Right}: Using
%$K_{2}$.}}{\Qlb{102}}{Figure}{\special{ language "Scientific Word";
%type "GRAPHIC";  maintain-aspect-ratio TRUE;  display "USEDEF";
%valid_file "T";  width 7.1018in;  height 2.2208in;  depth 0pt;
%original-width 9.1912in;  original-height 2.8539in;  cropleft "0";
%tempfilename 'POIP3I03.wmf';tempfile-properties "XPR";}} }%
%BeginExpansion
\begin{figure}
[ptb]
\begin{center}
\includegraphics[
%natheight=2.853900in,
%natwidth=9.191200in,
%height=2.2208in,
%width=7.1018in
scale=.4]{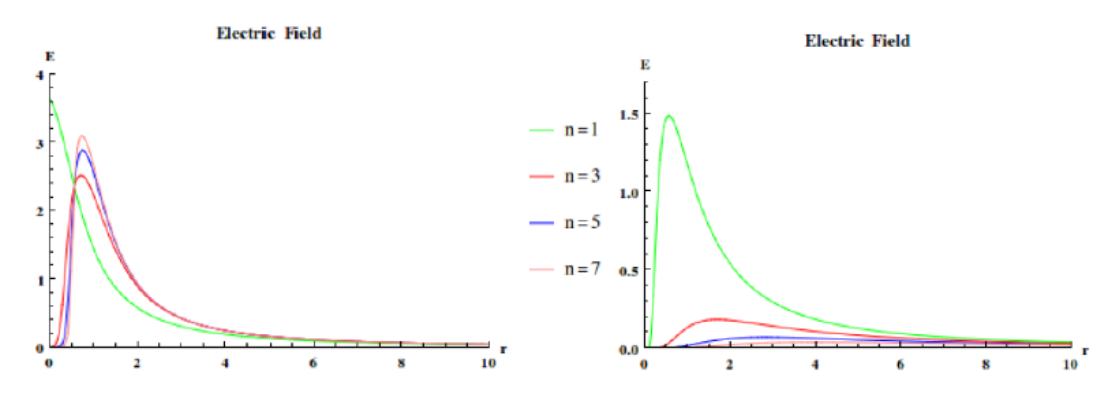}%
%{POIP3I03.wmf}%
\caption{Electric field for $n=1,3,5,7$. (a) \textbf{Left}: Using $K_{1}$. (b)
\textbf{Right}: Using $K_{2}$.}%
\label{102}%
\end{center}
\end{figure}
%EndExpansion
We can see the same behaviour in the electric and magnetic field,
Fig.\ref{102} and Fig.\ref{103} comparing $K_{1}$ (left figures) with $K_{2}$
(right figures). In Fig.\ref{104} for the density is analogous.\newline%
%TCIMACRO{\FRAME{ftbpFU}{5.38in}{1.8758in}{0pt}{\Qcb{Electric field for
%$n=1,3,5,7$. (a) \QTR{bf}{Left}: Using $K_{1}$. (b) \QTR{bf}{Right}: Using
%$K_{2}$.}}{\Qlb{103}}{Figure}{\special{ language "Scientific Word";
%type "GRAPHIC";  maintain-aspect-ratio TRUE;  display "USEDEF";
%valid_file "T";  width 5.38in;  height 1.8758in;  depth 0pt;
%original-width 9.0321in;  original-height 3.1332in;  cropleft "0";
%tempfilename 'POIP5O04.wmf';tempfile-properties "XPR";}} }%
%BeginExpansion
%EndExpansion
~
%TCIMACRO{\FRAME{ftbpFU}{6.026in}{1.9597in}{0pt}{\Qcb{Density for $n=1,3,5,7$.
%(a) \QTR{bf}{Left}: Using $K_{1}$. (b) \QTR{bf}{Right}: Using $K_{2}$%
%.}}{\Qlb{104}}{Figure}{\special{ language "Scientific Word";  type "GRAPHIC";
%maintain-aspect-ratio TRUE;  display "USEDEF";  valid_file "T";
%width 6.026in;  height 1.9597in;  depth 0pt;  original-width 8.7943in;
%original-height 2.8435in;  cropleft "0";  croptop "1";  cropright "1";
%cropbottom "0";  tempfilename 'POIP8H05.wmf';tempfile-properties "XPR";}} }%
%BeginExpansion
%EndExpansion
Respect to the accretion velocity, for $K_{1}$ the case $n=1$ is different to
the case using $K_{2}$ where all profiles has the same shape with different
intensities as shown in Fig.\ref{105}%
%TCIMACRO{\FRAME{ftbpFU}{5.585in}{1.7772in}{0pt}{\Qcb{Accretion velocity for
%$n=1,3,5,7$. (a) \QTR{bf}{Left}: Using $K_{1}$. (b) \QTR{bf}{Right}: Using
%$K_{2}$.}}{\Qlb{105}}{Figure}{\special{ language "Scientific Word";
%type "GRAPHIC";  maintain-aspect-ratio TRUE;  display "USEDEF";
%valid_file "T";  width 5.585in;  height 1.7772in;  depth 0pt;
%original-width 8.873in;  original-height 2.8046in;  cropleft "0";
%tempfilename 'POIPA706.wmf';tempfile-properties "XPR";}} }%
%BeginExpansion
%EndExpansion
%TCIMACRO{\FRAME{ftbpFU}{6.0597in}{1.9735in}{0pt}{\Qcb{Accretion rate for
%$n=1,3,5,7$. (a) \QTR{bf}{Left}: Using $K_{1}$. (b) \QTR{bf}{Right}: Using
%$K_{2}$.}}{\Qlb{106}}{Figure}{\special{ language "Scientific Word";
%type "GRAPHIC";  maintain-aspect-ratio TRUE;  display "USEDEF";
%valid_file "T";  width 6.0597in;  height 1.9735in;  depth 0pt;
%original-width 9.1912in;  original-height 2.9741in;  cropleft "0";
%tempfilename 'POIPBO07.wmf';tempfile-properties "XPR";}} }%
%BeginExpansion
\begin{figure}
[ptb]
\begin{center}
\includegraphics[
%natheight=2.974100in,
%natwidth=9.191200in,
%height=1.9735in,
%width=6.0597in
scale=.5]{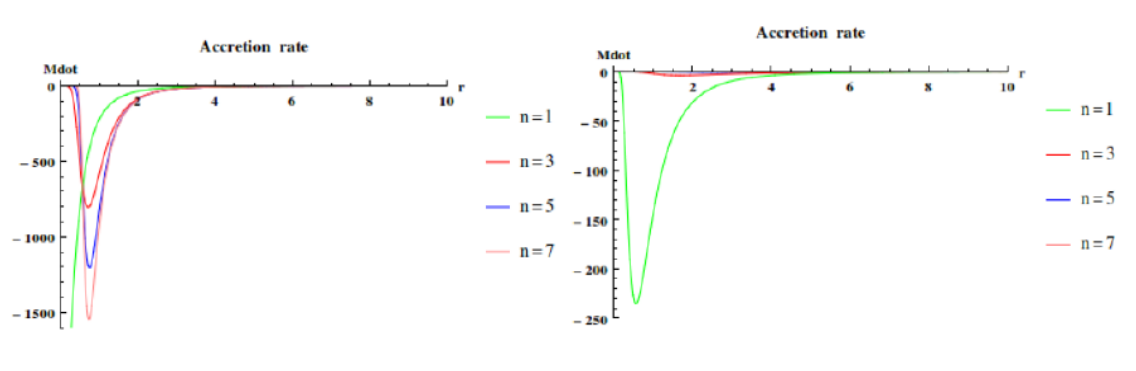}%
%{POIPBO07.wmf}%
\caption{Accretion rate for $n=1,3,5,7$. (a) \textbf{Left}: Using $K_{1}$. (b)
\textbf{Right}: Using $K_{2}$.}%
\label{106}%
\end{center}
\end{figure}
%EndExpansion

All the curves present qualitatively the same behaviour except the case $n=1$
using $K_{1}$, this case present accretion for the whole range of $r$,
meanwhile the another velocities goes to zero when $r\rightarrow0$. For
$n\neq1$ due to accretion velocity goes to zero could be interpreted as matter
co-rotating near the central object in $r=0$. For $K_{2}$ disappears the
divergence for $n=1$ as we expected. In Fig.\ref{106} the accretion mass rate
is completely consistent with the fact that is related to accretion velocity,
where the values changes in the velocity profiles according the changes in the
curvature.

\subsection{Introducing matter: Comparison for different values of $\omega$}

\begin{figure}
[ptb]
\begin{center}
\includegraphics[
%natheight=2.864300in,
%natwidth=8.953400in,
%height=1.5627in,
%width=4.8516in
scale=.55]{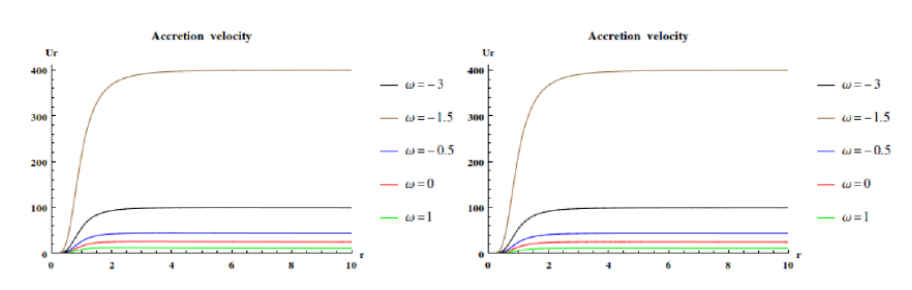}%
%{POIPCT08.wmf}%
\caption{Accretion velocity for different values of $\omega$ \textbf{Left}:
Using $K_{1}$. \textbf{Right}: Using $K_{2}$.}%
\label{107}%
\end{center}
\end{figure}

\begin{figure}
[ptb]
\begin{center}
\includegraphics[
%natheight=2.585800in,
%natwidth=8.397300in,
%height=1.4961in,
%width=4.8248in
scale=.5]{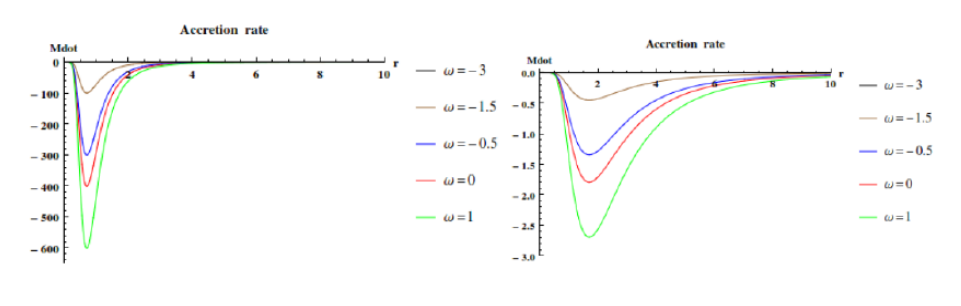}%
%{POIPDZ09.wmf}%
\caption{Accretion rate for different values of $\omega$ \textbf{Left}: Using
$K_{1}$. \textbf{Right}: Using $K_{2}$.}%
\label{108}%
\end{center}
\end{figure}

We had set: $A_{4}=10$, $A_{2}=1$, $m=1$, $M=1$, $n=3$, $e=1$, $b=3.8296591$,
$A_{0}=2+\omega$ and $a=-0.9$ for all profiles. In order to compare with
\cite{BJ15}we choose different values of $\omega$ as we can see in
Fig.\ref{107} and Fig.\ref{108}.%
%TCIMACRO{\FRAME{ftbpFU}{4.8516in}{1.5627in}{0pt}{\Qcb{Accretion velocity for
%different values of $\omega$ \QTR{bf}{Left}: Using $K_{1}$. \QTR{bf}{Right}:
%Using $K_{2}$.}}{\Qlb{107}}{Figure}{\special{ language "Scientific Word";
%type "GRAPHIC";  maintain-aspect-ratio TRUE;  display "USEDEF";
%valid_file "T";  width 4.8516in;  height 1.5627in;  depth 0pt;
%original-width 8.9534in;  original-height 2.8643in;  cropleft "0";
%tempfilename 'POIPCT08.wmf';tempfile-properties "XPR";}} }%
%BeginExpansion
%EndExpansion
The cases where $\omega<0$ corresponds to Dark Matter. The values are chosen
in order to compare with \cite{BJ15}. In particular, velocity profiles are far
from Keplerian profiles as we can expect for galaxy models including Dark
Matter qualitatively coherent with observational profiles. Also, as we can see
in Fig.\ref{108} both kernels, $K_{1}$ and $K_{2}$ gives different accretion
rates, however we have similar accretion velocities.%
%TCIMACRO{\FRAME{ftbpFU}{4.8248in}{1.4961in}{0pt}{\Qcb{Accretion rate for
%different values of $\omega$ \QTR{bf}{Left}: Using $K_{1}$. \QTR{bf}{Right}:
%Using $K_{2}$.}}{\Qlb{108}}{Figure}{\special{ language "Scientific Word";
%type "GRAPHIC";  maintain-aspect-ratio TRUE;  display "USEDEF";
%valid_file "T";  width 4.8248in;  height 1.4961in;  depth 0pt;
%original-width 8.3973in;  original-height 2.5858in;  cropleft "0";
%tempfilename 'POIPDZ09.wmf';tempfile-properties "XPR";}} }%
%BeginExpansion
%EndExpansion
~

\subsection{BPS\ condition: the Reissner-Nordstr\"{o}m limit}

In this section we had set: $A_{4}=10$, $A_{2}=1$, $m=1$, $M=1$, $n=3$, $e=1$,
$b=3.8296591$ $\omega=0$ and $a=-0.9$ for all profiles.%
%TCIMACRO{\FRAME{ftbpFU}{4.9398in}{1.4762in}{0pt}{\Qcb{Metric coefficient
%$A\left(  r\right)  $ for $\ n=1,3,5,7.$ \QTR{bf}{Left:} Using $K_{1}.$
%\QTR{bf}{Right:} Using $K_{2}$.}}{\Qlb{109}}{Figure}%
%maintain-aspect-ratio TRUE;  display "USEDEF";  valid_file "T";
%width 4.9398in;  height 1.4762in;  depth 0pt;  original-width 9.3097in;
%original-height 2.7648in;  cropleft "0";  croptop "1";  cropright "1";
%cropbottom "0";  tempfilename 'POIPFA0A.wmf';tempfile-properties "XPR";}} }%
%BeginExpansion
\begin{figure}
[ptb]
\begin{center}
\includegraphics[
%natheight=2.764800in,
%natwidth=9.309700in,
%height=1.4762in,
%width=4.9398in
scale=.5]{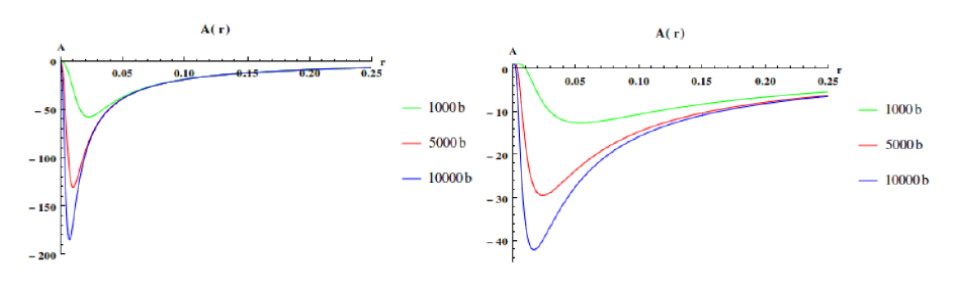}%
%{POIPFA0A.wmf}%
\caption{Metric coefficient $A\left(  r\right)  $ for $\ n=1,3,5,7.$
\textbf{Left:} Using $K_{1}.$ \textbf{Right:} Using $K_{2}$.}%
\label{109}%
\end{center}
\end{figure}
%EndExpansion
%TCIMACRO{\FRAME{ftbpFU}{4.7115in}{1.3318in}{0pt}{\Qcb{Density for
%$\ n=1,3,5,7.$ \QTR{bf}{Left:} Using $K_{1}.$ \QTR{bf}{Right:} Using $K_{2}%
%$.}}{\Qlb{110}}{Figure}{\special{ language "Scientific Word";
%type "GRAPHIC";  maintain-aspect-ratio TRUE;  display "USEDEF";
%valid_file "T";  width 4.7115in;  height 1.3318in;  depth 0pt;
%original-width 8.8341in;  original-height 2.4768in;  cropleft "0";
%tempfilename 'POIPGY0B.wmf';tempfile-properties "XPR";}} }%
%BeginExpansion
\begin{figure}
[ptb]
\begin{center}
\includegraphics[
%natheight=2.476800in,
%natwidth=8.834100in,
%height=1.3318in,
%width=4.7115in
scale=.5]{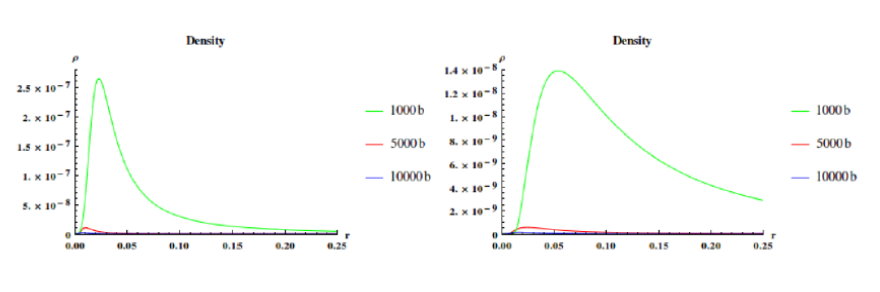}%
%{POIPGY0B.wmf}%
\caption{Density for $\ n=1,3,5,7.$ \textbf{Left:} Using $K_{1}.$
\textbf{Right:} Using $K_{2}$.}%
\label{110}%
\end{center}
\end{figure}
%EndExpansion
%TCIMACRO{\FRAME{ftbpFU}{4.7937in}{1.369in}{0pt}{\Qcb{Accretion velocity for
%$\ n=1,3,5,7.$ \QTR{bf}{Left:} Using $K_{1}.$ \QTR{bf}{Right:} Using $K_{2}%
%$.}}{\Qlb{111}}{Figure}{\special{ language "Scientific Word";
%type "GRAPHIC";  maintain-aspect-ratio TRUE;  display "USEDEF";
%valid_file "T";  width 4.7937in;  height 1.369in;  depth 0pt;
%original-width 9.5475in;  original-height 2.7051in;  cropleft "0";
%tempfilename 'POIPI10C.wmf';tempfile-properties "XPR";}} }%
%BeginExpansion
\begin{figure}
[ptb]
\begin{center}
\includegraphics[
%natheight=2.705100in,
%natwidth=9.547500in,
%height=1.369in,
%width=4.7937in
scale=.5]{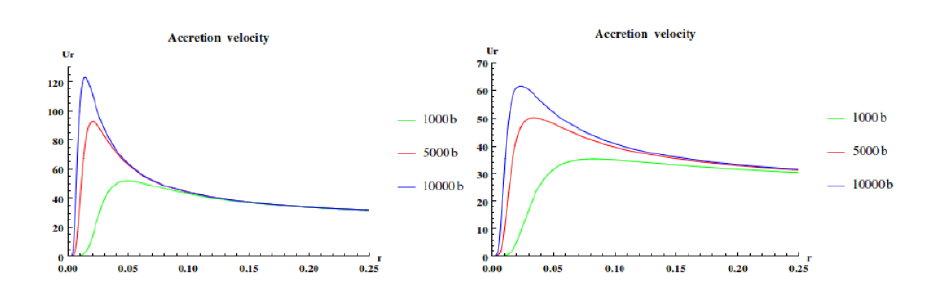}%
%{POIPI10C.wmf}%
\caption{Accretion velocity for $\ n=1,3,5,7.$ \textbf{Left:} Using $K_{1}.$
\textbf{Right:} Using $K_{2}$.}%
\label{111}%
\end{center}
\end{figure}
%EndExpansion
%TCIMACRO{\FRAME{ftbpFU}{4.6371in}{1.3188in}{0pt}{\Qcb{Accretion rate for
%$\ n=1,3,5,7.$ \QTR{bf}{Left:} Using $K_{1}.$ \QTR{bf}{Right:} Using $K_{2}%
%$.}}{\Qlb{112}}{Figure}{\special{ language "Scientific Word";
%type "GRAPHIC";  maintain-aspect-ratio TRUE;  display "USEDEF";
%valid_file "T";  width 4.6371in;  height 1.3188in;  depth 0pt;
%original-width 9.5475in;  original-height 2.6956in;  cropleft "0";
%tempfilename 'POIPJG0D.wmf';tempfile-properties "XPR";}} }%
%BeginExpansion
\begin{figure}
[ptb]
\begin{center}
\includegraphics[
%natheight=2.695600in,
%natwidth=9.547500in,
%height=1.3188in,
%width=4.6371in
scale=.5]{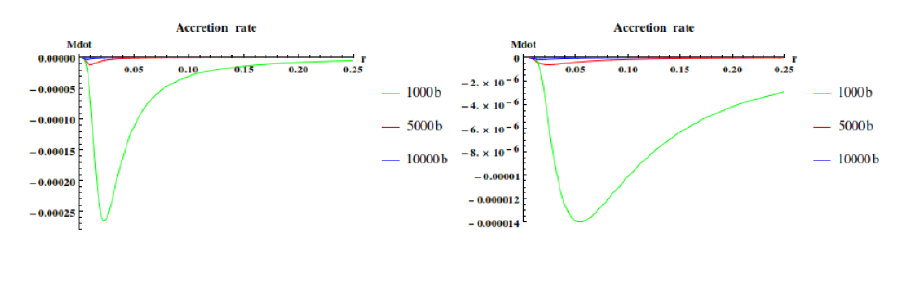}%
%{POIPJG0D.wmf}%
\caption{Accretion rate for $\ n=1,3,5,7.$ \textbf{Left:} Using $K_{1}.$
\textbf{Right:} Using $K_{2}$.}%
\label{112}%
\end{center}
\end{figure}
%EndExpansion
As it is known the action of EBI is the basis of the theories of great
unification in particular theories containing supersymmetries as a fundamental
ingredient. All these theories to be endowed with a high geometric content,
such as theory of membranes, matrices and superstrings, which contain a
non-linear non-commutative structure and therefore not local. In such theories
the minimum energy condition or BPS corresponds to the linear or maxwellian
limit, which in the case of Spherical Symmetric Solution (SSS) in theories of
EBI are transformed into the solution of Reissner-Nordstr\"{o}m (RN) limit.
The importance from the astrophysical point of view that we propose here is
that, when the condition of regularity is broken that is when one goes to the
linear limit, the singularities appear generating the divergences that affect
the mechanisms of stability and accretion.

In the case of unified theories as in the case of MacDowell-Mansouri 1979
\cite{manso} and Cirilo-Lombardo 2017 \cite{diego17}, the limit value of the
fields of the theory (which plays the role of the arbitrary parameter b in the
standard EBI theory) is subject to the curvature and the dynamics of the same
physical fields (states) that intervene both in the accretion and in the
structure of space-time. This is completely in accordance with the conjecture
in MacDowell-Mansouri 1979 and completely tested in Cirilo-Lombardo 2017.

\section{Concluding remarks}

In this work, we have been involved in the construction of gravastar models
inspired in the full exact regular solutions of EBI nonlinear electrodynamics
given in \cite{DCL} and a new one of the Yukawa type. Although in this paper
the back-reaction is not included, as is known, by means of a perturbative
treatment, the back-reaction can be included in a simple manner with a minimum
of modifications in both the Einstein equations and the accretion ones.
Spherically symmetric solutions with a regular center (as our case) are stable
before disturbances\cite{br}, consequently the methods of treating the
back-reaction are always favored in these scenarios Therefore, due to the full
regularity in the gravastar models that we constructed, it is not necessary to
regain the trouble with all type of spacetime singularities when accretion
mechanisms are considered.

We use the proposed framework for the study of spherical accretion onto
spherically symmetric compact objects given by Bahamonde and Jamil
\cite{BJ15}. Despite having been indicated by those authors that the context
of their proposal is the most general for spherically symmetric
configurations, this proposal is only a rough approximation to the true
problem of accretion of compact objects where time-dependent solutions are
also ne\-ce\-ssa\-ry for a realistic description of the astrophysical process.
We explored various features of our gravastar type solutions in the Bahamonde
and Jamil framework considering no matter: \textit{e.g.} only the
selfgravitating nonlinear electromagmnetic configuration and with matter
subject to the simplest state equation of type $p(r)=(1+\omega)\rho(r)$. This
equation allows to see the behavior, in particular the stability of the
solution before the accretion of normal matter and the exotic matter (dark
matter) with respect to the values of the parameters chosen for both the
particular solution of Cirilo-Lombardo (2005) \cite{DCL} and the new solution
of exponential type (Yukawa).

Concerning the accretion velocity profiles we can see that in the cases with
exponential (Yukawa type ) solutions the shape of the curves are more smooth
that in the case of the non exponential one \cite{DCL}. Consequently we can
expect that the inclusion in any galactic model of a regular EBI field can
explain the several discrepancies between Modified Newtonian Dynamics (MOND)
and pure dark matter proposals. If well in the actual research apparently when
two modifications of the galaxy rotation curves from both MOND and the
addition of Dark Matter change the curves in the desired way solving the
troubles (by hand), this modification is only a static solution that works on
simple systems. When applied to a more complex system the standard
modifications does not solve the entire problem of the mass discrepancy and
only manages to reduce the difference. The solution that Dark Matter provides
is to add unobservable additional matter to the mass distribution. This idea
of extra unseen mass is also supported by gravitational lensing experiments.
There have been a number of propositions for the shape of the dark matter
distribution, since this is unknown. These different proposals all lead to
similar rotation curves with minor differences. Some rise quicker and decrease
faster, like the Einasto-B profile, others rise slower but flatten off instead
of decreasing, like the Burkert profile. Then, we remark here that in our case
precisely the electromagnetic field through the density obtained from the same
energy-momentum tensor, is able to show similar profiles to those considered
empirically in galactic models of dark matter. Therefore the EBI
electromagnetic fields in the context of regular spherically symmetric
solutions must be considered together with the other proposals of both MOND
and dark matter to solve all the discrepancies.

\section{Acknowledgements}

We gratefully acknowledge to the Departamento de F\'{\i}sica (FCEyN, UBA) and
Instituto de F\'{\i}sica del Plasma (CONICET-UBA), DC is also grateful to the
Consejo Nacional de Investigaciones Cient\'{\i}ficas y T\'{e}cnicas and to the
Bogoliubov Laboratory of Theoretical Physics, CV is also grateful to the
Instituto de Ciencias (UNGS) for their institutional support.

\bibliographystyle{plain}
\bibliography{biblio}

\end{document}